\documentclass[doublecol]{epl2} 
\usepackage{graphicx}
\usepackage{amsfonts}
\usepackage{bm}
\usepackage{amsmath,amssymb,float}

\newcommand* {\ee}{\ensuremath{\mathrm{e}}}

\title{Neutral edge modes in a superconductor -- topological-insulator
  hybrid structure in a perpendicular magnetic field}

\author{Rakesh P. Tiwari\inst{1} \and U. Z\"ulicke\inst{2} \and C. Bruder\inst{1} \and Vladimir M. Stojanovi\'c\inst{3}}
\shortauthor{Rakesh P. Tiwari \etal.}

\institute{                    
  \inst{1} Department of Physics, University of Basel,
  Klingelbergstrasse 82, CH-4056 Basel, Switzerland\\
  \inst{2} School of Chemical and Physical Sciences and MacDiarmid
  Institute for Advanced Materials and Nanotechnology, Victoria
  University of Wellington, PO Box 600, Wellington 6140, New Zealand\\
  \inst{3} Department of Physics, Harvard University, Cambridge, MA
  02138, USA
}

\abstract{
  We study the low-energy edge states of a superconductor -- 3D
  topological-insulator hybrid structure (NS junction) in the presence
  of a perpendicular magnetic field. The hybridization of
  electron-like and hole-like Landau levels due to Andreev reflection
  gives rise to chiral edge states within each Landau level. We show
  that by changing the chemical potential of the superconductor, this
  junction can be placed in a regime where the sign of the effective
  charge of the edge state within the zeroth Landau level changes more
  than once resulting in \textit{neutral} edge modes with a finite
  value of the guiding-center coordinate.  The appearance
  of these neutral edge modes is related to the level repulsion
  between the zeroth and the first Landau levels in the spectra. We
  also find that these neutral edge modes come in \textit{pairs}, one
  in the zeroth Landau level and its corresponding pair in the first.
  Unlike ordinary neutral \textit{bogolon\/} excitations in
  superconductors, the neutral modes found by us have a finite speed
  and, thus, the potential to carry a heat current.}

\begin{document}

\maketitle

\section{Introduction}
The integer quantum Hall (QH) state~\cite{Klitzing+:80} represented
the first example of a many-body state whose nontrivial properties do
not stem from a spontaneously-broken symmetry, but rather its
topological character~\cite{Girvin+Prange:90}. The precise
quantisation of the Hall conductance in integer multiples of $e^{2}/h$
was initially studied using a two-dimensional electron gas at a low
temperature and in a strong perpendicular magnetic field, receiving a
renewed attention with the discovery of graphene and its peculiar
Dirac-like energy spectrum~\cite{Zhang+:05,Abanin+:06}.  At the heart
of transport in integer-QH systems -- which based on their energy
spectra could be expected to be insulating if the chemical potential
lies in the gap between the neighbouring Landau levels (LLs) -- is the
existence of edge states. These LLs acquire dispersion as the guiding
centers approach the physical edges of the sample (Hall bar), which
gives rise to a current along the edges. The net current, resulting
from a voltage (or chemical potential) drop in the direction
perpendicular to the Hall-bar edges, is determined by the number of
edge channels, {\it i.e.}, of LLs occupied in the bulk.

The advent of quantum spin-Hall
states~\cite{Murakami+:03,Kane+Mele:05}, topologically distinct from
all previously known states of matter, led to a resurgence of interest
in edge states. In this particular context, the peculiar feature of
the ensuing topological insulators (TIs)~\cite{Bernevig+:06} is that
edge-state transport in these bulk-insulating materials takes place in
a time-reversal-invariant fashion, {\it i.e.}, without an external magnetic
field (see Refs.~\cite{TIreviews1,TIreviews2}).  This is their
principal difference from the conventional edge states in integer-QH
systems.

Apart from integer-QH and TI systems, edge-state physics has in recent
years been studied in heterostructures. The prime examples are the
normal-superconductor (NS) junctions in a perpendicular magnetic
field, whose N-part entails a system with a Dirac-like energy spectrum
(either graphene or a TI).  An additional physical ingredient in these
systems -- compared to the two aforementioned instances of edge states
-- is Andreev reflection~\cite{Andreev:64} at the NS interface. An
electron (charge $-e$) incident from the N side is reflected as a
positively-charged hole, while the missing charge of $-2e$ enters the
superconductor as an electron pair. These electron-hole conversion
processes~\cite{Beenakker:06} and the ensuing Andreev edge states have
been studied both theoretically~\cite{akh07} and
experimentally~\cite{ric12} in graphene contacted with superconducting
electrodes.
\begin{figure}
\includegraphics[width=.925\columnwidth]{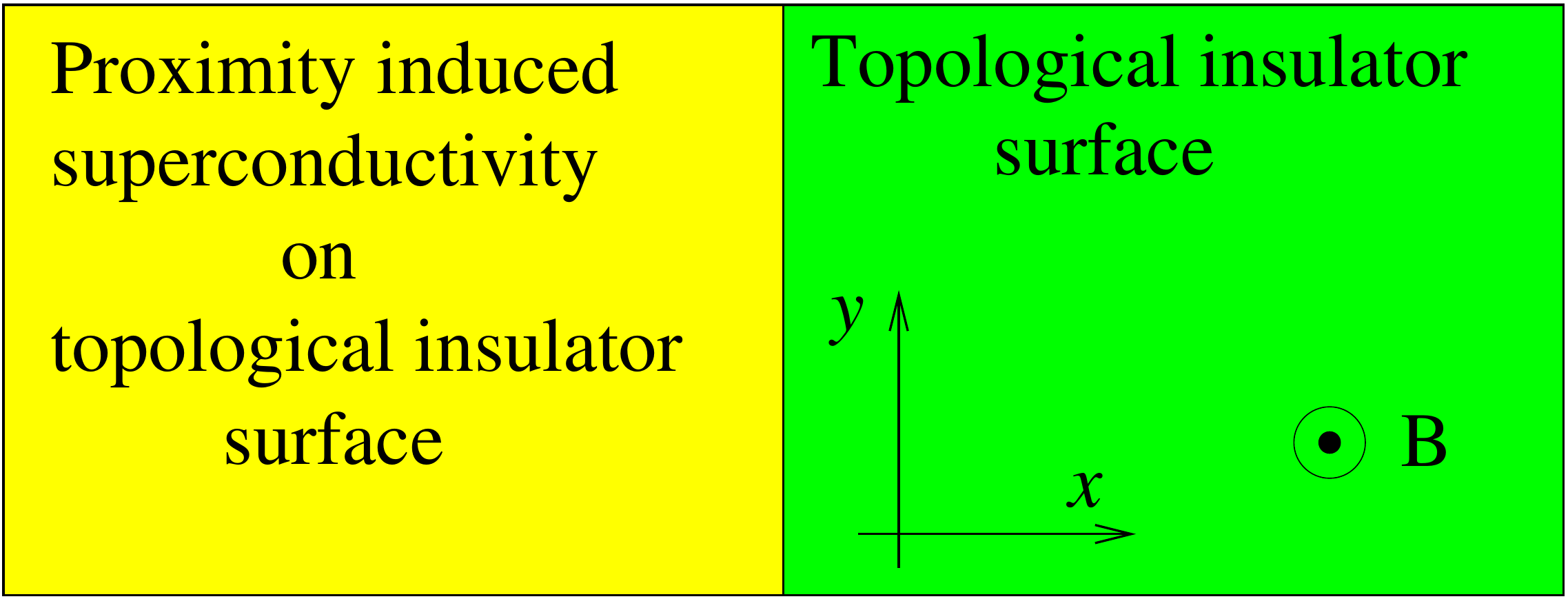}
\caption{\label{fig:schematic}(Colour online) {Schematic of the NS
  heterostructure based upon a 3D topological insulator (TI). The left
  half represents the proximity-induced superconducting surface states
  of the TI, while the right one represents the normal surface states
  of the TI. A perpendicular magnetic field is
  applied in the right half.}}
\end{figure}

Quite recently, it was shown that a superconductor-TI junction in the
presence of a perpendicular magnetic field can support a neutral
chiral Majorana mode within the zeroth LL~\cite{tiw13}, which exists
when the value of the guiding-center coordinate is \textit{zero}.
In contrast, in this article, we investigate edge states with a
finite value of the guiding-center coordinate.  We show that this
junction supports additional \textit{neutral} edge modes within the
zeroth LL, which arise due to level repulsion between the zeroth and
the first LLs in the spectrum.  Using the Dirac-Bogoliubov-de~Gennes
equation~\cite{Beenakker:06,akh07,fu08}, we study the spectrum of the
dispersive edge states and show that this spectrum depends strongly on
the chemical potential of the superconductor. We predict the existence
of finite-energy, propagating neutral fermionic edge channels
for a nonzero value of the guiding-center coordinate when the chemical
potential of the superconductor is larger than 
a critical value $\mu_c$. These edge modes do not obey the
self-conjugation property, which makes them fundamentally different
from the Majorana modes studied in ref.~\cite{tiw13}.


\section{System and model}
\label{sec:system}
\begin{figure}
\includegraphics[width=.99\columnwidth,clip=true]{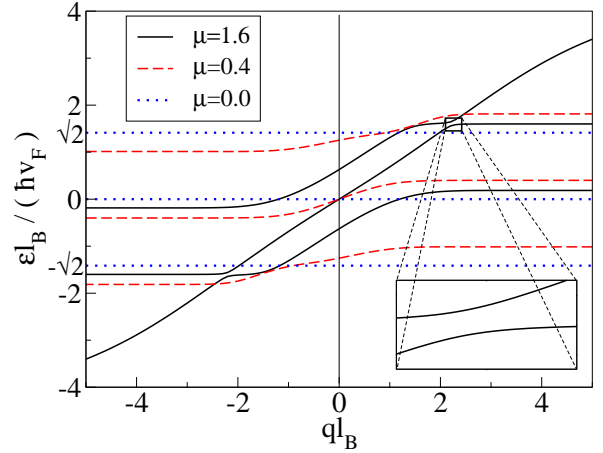}
\caption{\label{fig:dispersion} (Colour online) 
 Energy dispersion relation $\varepsilon_{n}(q)$, showing the zeroth
 and the first electron and hole-like ($n=\pm1$) Landau levels (LLs). 
 Dotted (blue) lines are for $\mu=0$, dashed (red) 
 lines for $\mu=0.4 \hbar v_F/l_B$, and solid (black) lines for 
 $\mu=1.6 \hbar v_F/l_B$. At $q=0$, for all values of $\mu$, the
 zeroth LL has zero energy, the $n=1$ LL 
  has positive energy, and the $n=-1$ LL has negative energy. 
  The value of the superconducting gap is $\Delta_0=5 \hbar v_F/l_B$.
The inset shows a zoom-in of the level repulsion between the zeroth
and first LLs for $\mu=1.6\hbar v_F/l_B$.
}
\end{figure}
The system under investigation is shown in fig.~\ref{fig:schematic}.
A conventional superconductor is deposited on the top surface of a 3D
TI (depicted as $x-y$ plane in fig.~\ref{fig:schematic}) inducing a
finite superconducting pairing amplitude $\Delta({\bf
  r})=\ee^{i\theta}\Delta_0$ in the left half ($x<0$) via proximity
effect. A finite magnetic field ${\bf B}={\bf \nabla}\times{\bf A}$ is
present in the right half ($x>0$), where ${\bf A}$ denotes the vector
potential.  The single-particle excitations in this N--S
heterostructure can be described by the Dirac-Bogoliubov-de~Gennes
equation~\cite{Beenakker:06,akh07,fu08,tiw13}
\begin{equation}
\left( \begin{array}{cc}
H_{D}({\bf r})-\mu & \Delta({\bf r})\,\sigma_0 \\ 
\Delta^{\ast}({\bf r})\, \sigma_0 &
\mu-\mathcal{T} H_{D}({\bf r})\, \mathcal{T}^{-1} \end{array}\right) 
\Psi({\bf r}) = \varepsilon \Psi({\bf r}) \:.
\label{eq:hamiltonian} 
\end{equation}
Here $\sigma_0\equiv\mathbb{I}_{2\times 2}$ denotes the
two-dimensional identity matrix, $\mathcal{T}$ the time-reversal
operator, and $H_{D}({\bf r}) = v_{F} [{\bf p} + e{\bf A}({\bf r})]
\cdot \bm{\sigma}$ the massless-Dirac Hamiltonian for the TI surface
states, with $\bm{\sigma}$ being the vector of Pauli matrices that act
in spin space. The position ${\bf r}\equiv (x, y)$ and momentum ${\bf
  p}\equiv -i\hbar(\partial_x, \partial_y)$ are restricted to the TI
surface. In eq.~(\ref{eq:hamiltonian}) the excitation energy
$\varepsilon$ is measured relative to the chemical potential $\mu$ of
the superconductor, with the absolute zero of this energy set to be at
the Dirac ({\it i.e.}, charge neutrality) point of the TI surface
states. The wave function $\Psi=(u_{\uparrow},
u_{\downarrow},v_{\downarrow},-v_{\uparrow})^{T}$ in
eq.~(\ref{eq:hamiltonian}) is a spinor in Dirac-Nambu space.  Choosing
the Landau gauge ${\bf A}=B\, x\, \hat{{\bf y}}$ and demanding
continuity of the wavefunction across the interface, we numerically
obtain the energy-dispersion relation $\varepsilon=\varepsilon(q)$,
where $q$ is the guiding-center coordinate.  The
explicit form of the Dirac-Nambu spinor in the normal region is
\begin{align}
&\Psi(x,y)= e^{iqy}\nonumber\\
\times&\left(\begin{array}{c}
-i C_e e^{-(x + q)^2/2}(\mu +\varepsilon)H_{(\mu +\varepsilon)^2/2 -1}(x +q) e^{i\theta/2} \\
C_e e^{-(x + q)^2/2} H_{(\mu +\varepsilon)^2/2}(x +q) e^{i\theta/2} \\
C_h e^{-(x - q)^2/2} H_{(\mu -\varepsilon)^2/2}(x -q) e^{-i\theta/2} \\
-i C_h e^{-(x - q)^2/2}(\mu -\varepsilon)H_{(\mu -\varepsilon)^2/2 -1}(x -q) e^{-i\theta/2}
\end{array}
\right),
\label{eq:wfn}
\end{align}
where $H_{\alpha}(x)$ stands for the Hermite
function~\cite{akh07,tiw13}. The coefficients $C_e$ and $C_h$ satisfy
\begin{align}
\frac{C_e}{C_h}=\frac{-i\Delta_0(\mu-\varepsilon)H_{(\mu
    -\varepsilon)^2/2 -1}(-q)}{\varepsilon H_{(\mu +\varepsilon)^2/2}(q) +(\mu +\varepsilon)H_{(\mu +\varepsilon)^2/2 -1}(q) \sqrt{\Delta_0^2-\varepsilon^2}}. 
\end{align}

All lengths are measured in units of the
magnetic length $l_B\equiv\sqrt{\hbar/|eB|}$ and energies in units of
$\hbar v_F/l_B$; the dimensionless parameter representing the guiding
center coordinate is $ql_B$.

\section{Results}
\label{sec:results}
Figure~\ref{fig:dispersion} shows the dispersion relation
$\varepsilon_{n}(q)$ of the single-particle excitations, where $n$
denotes the LL index.  Only the zeroth ($n=0$) and the first LLs
($n=\pm1$) are shown for $\mu=0$ (dotted lines), $\mu=0.4\hbar
v_F/l_B$ (dashed lines), and $\mu=1.6\hbar v_F/l_B$ (solid lines).
The
dispersion relation has inversion symmetry
$\varepsilon_{n}(q)=-\varepsilon_{-n}(-q)$.  This energy spectrum
contains a chiral Majorana mode within the zeroth Landau level around
$q=0$~\cite{tiw13}.  In the present article we focus on the level
repulsion between different Landau levels at {\em finite}
$q$. Figure~\ref{fig:dispersion} shows these level repulsions between
the zeroth ($n=0$) and the first ($n=1$) LLs for $\mu=1.6 \hbar v_F/l_B$
and a similar one between the zeroth ($n=0$) and the first hole-like
($n=-1$) LLs. It should be noted that there is no such level repulsion
for $\mu=0$ and $\mu={0.4} \hbar v_F/l_B$.

\begin{figure}
\includegraphics[height=1.0\columnwidth,clip=true,angle=270]{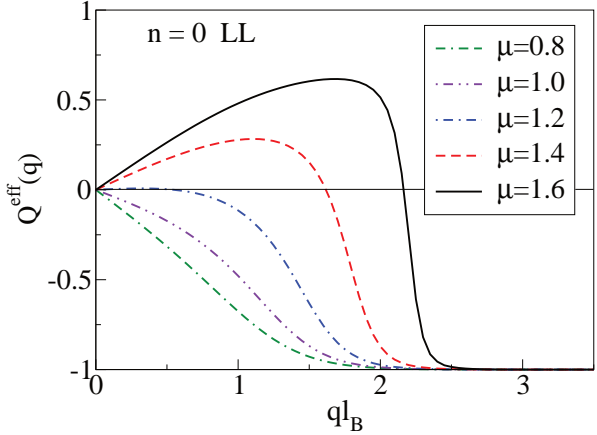}
\caption{\label{fig:effCharge} (Colour online) Effective Nambu charge
  $Q^{\text{eff}}(q)$ calculated using eq.~(\ref{eq:qeff}) for different
  values of $\mu$ (expressed in units of $\hbar v_F/l_B$), for the
  zeroth LL.}
\end{figure}
The effective Nambu charge 
\begin{equation}
Q^{\text{eff}}(q)=\int_{0}^{\infty} d x \, \left[\Psi(x,y)\right]^\dagger 
\sigma_0\otimes \tau_z \, \Psi(x,y)
\label{eq:qeff}
\end{equation} 
of the Dirac-Andreev edge states described by $\Psi(x,y)$ is generally
not quantised~\cite{deg89,kiv90,nay01,fuj08,par12} and can be
calculated from the solution of eq.~(\ref{eq:hamiltonian}). Here
$\tau_z$ represents the usual Pauli $z$ matrix that acts in the Nambu
space. We observe that $Q^{\text{eff}}(q)$ for the zeroth LL can
change sign more than once. It should be noted that one sign change at
$q=0$ is required as $Q^{\text{eff}}(q)$ vanishes linearly for
$ql_B\ll1$, due to the presence of a chiral Majorana mode within the
zeroth LL~\cite{tiw13}. We find that this additional change of sign
can be understood in terms of the level repulsion between the
neighbouring LLs. Figure~\ref{fig:effCharge} shows the numerically
computed $Q^{\text{eff}}(q)$ for the zeroth LL for various different
values of $\mu$. Only $q>0$ is shown in fig.~\ref{fig:effCharge} as
the effective charge for the zeroth LL has inversion symmetry,
$Q^{\text{eff}}(q)=-Q^{\text{eff}}(-q)$. For $\mu=0.8 \hbar v_F/l_B$
and $\mu=1.0 \hbar v_F/l_B$ we find that $Q^{\text{eff}}(q)$ remains
negative for all $q>0$, while for $\mu=1.4 \hbar v_F/l_B$ and $\mu=1.6
\hbar v_F/l_B$ it changes sign, indicating the presence of a
\textit{neutral} fermionic mode at a finite value of $q$. The presence
of this neutral fermionic mode has a direct correspondence with the
level repulsion between the zeroth and the first LLs for
$\mu\gtrsim1.2\hbar v_F/l_B$.

\begin{figure}[b!]
\includegraphics[width=.925\columnwidth,clip=true]{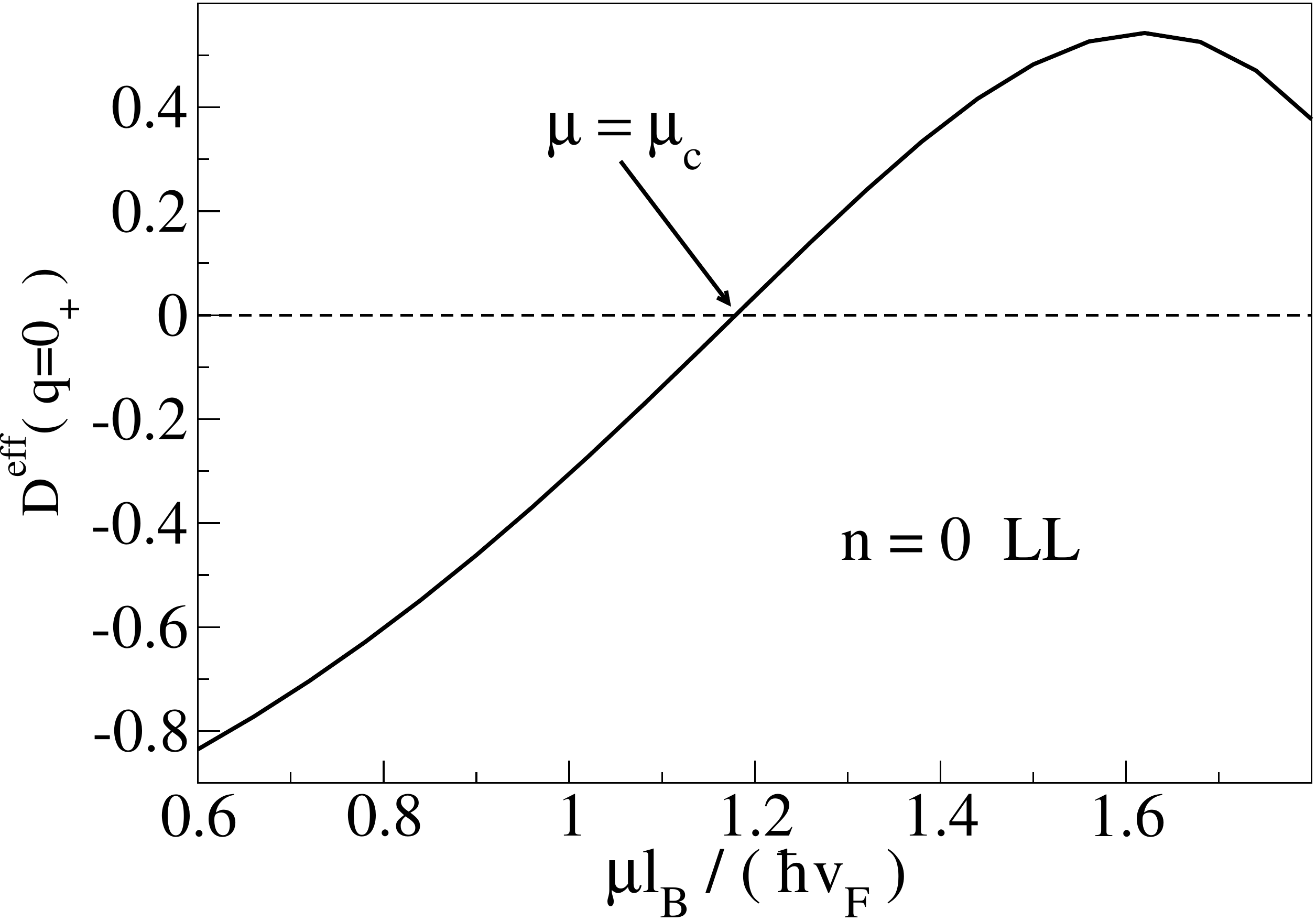}
\caption{\label{fig:dedqvmu} Derivative of the the effective Nambu
  charge $D^{\text{eff}}(q=0_{+})$
  as a function of the chemical potential $\mu$ for the zeroth LL.}
\end{figure}

For $\mu=0$, the dispersionless zeroth LL has a negative effective
charge over the entire domain $q>0$, signifying its
\textit{electron-like} character.  As $\mu$ is increased,  
the zeroth LL changes character from
\textit{electron-like} to \textit{hole-like} near the edge ($0 <
ql_B\lesssim 1$), while retaining its \textit{electron-like} character
deep inside the bulk ($ql_B\gg1$). This change necessitates the
presence of a neutral fermionic mode for a finite value of the
guiding-center coordinate $q\neq0$. 
Although not directly measurable, the derivative of the effective Nambu charge 
\begin{equation}
D^{\text{eff}}(q)=\frac{\partial Q^{\text{eff}}(q)}{\partial q}
\end{equation}
is a convenient quantity to discuss
the emergence of new neutral Andreev edge modes in our system.
\begin{figure}[b!]
\includegraphics[width=.925\columnwidth,clip=true]{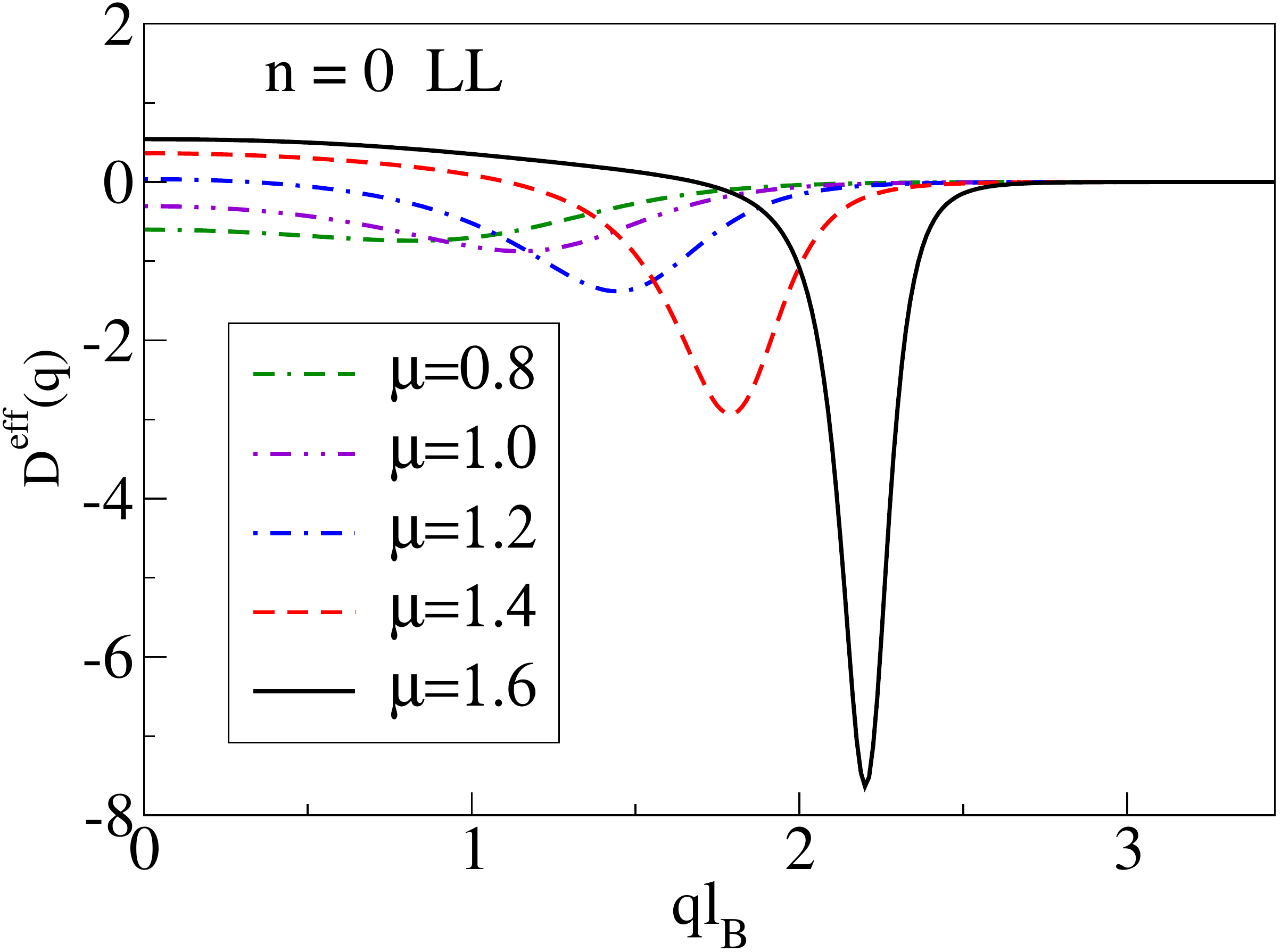}
\caption{\label{fig:dedqvsq} (Colour online) Derivative of
  the effective Nambu charge
  $D^{\text{eff}}(q)\equiv\partial Q^{\text{eff}}(q)/\partial q$ 
  for the zeroth LL and different values of
  $\mu$ (expressed in units of $\hbar v_F/l_B$).}
\end{figure}
Figure~\ref{fig:dedqvsq} shows a plot of $D^{\text{eff}}(q)$ for the
zeroth LL and different values of $\mu$.  We find that
$D^{\text{eff}}(q)$ remains a constant over most of its domain,
vanishing for $ql_B\gg1$ (independent of $\mu$). For $ql_B\sim1$,
$D^{\text{eff}}(q)$ shows a dip and finally approaches a constant
value again for $ql_B\ll1$.  The sign of the constant to which
$D^{\text{eff}}(q)$ approaches for $ql_B\ll1$ depends upon the value
of $\mu$. This neutral mode can be quantified further by calculating
$D^{\text{eff}}(q)$ at $q=0_{+}$. Figure~\ref{fig:dedqvmu} shows a
plot of $D^{\text{eff}}(q)$ calculated at $q=0_{+}$ as a function of
$\mu$. The critical value of the chemical potential $\mu_c$ is
indicated by the vanishing of $D^{\text{eff}}(q)$ at $q=0_{+}$.  For
$\mu > \mu_c$ a neutral fermionic edge mode exists within the zeroth
LL for a finite non-zero value of $q$.  As mentioned before, this
neutral fermionic mode is different from the chiral Majorana mode
discussed in ref.~\cite{tiw13}, which exists within the zeroth LL at
$q=0$ in that it does not satisfy self-conjugation criterion, a
defining property of the Majorana mode. The particle-hole conjugation
operator is $\Xi = \sigma _y \otimes \tau _y \mathcal{K}$, where
$\mathcal{K}$ denotes complex conjugation, $\sigma_y$ and $\tau_y$
are the Pauli $y$ matrices acting on spin and Nambu space
respectively. Therefore, the particle-hole conjugate of this neutral
mode at finite $q$, is another neutral mode at $-q$, also within the
zeroth LL.

\section{Discussion}
\label{sec:discussion}

At the interface of the NS junction, the superconducting region
hybridises the electron and the hole-like LLs resulting in dispersive
edge modes with non-quantised effective charge within the
superconducting gap. 
It should be noted that for $\mu=0$ the superconducting gap opens up
at the Dirac point such that the \textit{electron-like} branch of the
superconductor is located strictly above the Dirac point (conduction
band) while the \textit{hole-like} branch is located strictly below
the Dirac point (valence band). As $\mu$ increases, part of the
\textit{hole-like} branch of the superconductor starts belonging to
the conduction band. When $\mu > \sqrt{2} \hbar v_F/l_B$ the entire
$n=-1$ LL belongs to the conduction band, putting the spectrum in the
regime where level repulsion between the zeroth and the $n=1$ LL
results in a neutral fermionic edge mode at finite positive energy and
guiding-center coordinate. Similarly, level repulsion for the zeroth and
the $n=-1$ LL results in another neutral fermionic edge mode at
finite negative energy and guiding-center coordinate.  

\begin{figure}[h]
\includegraphics[width=.625\columnwidth,angle=270]{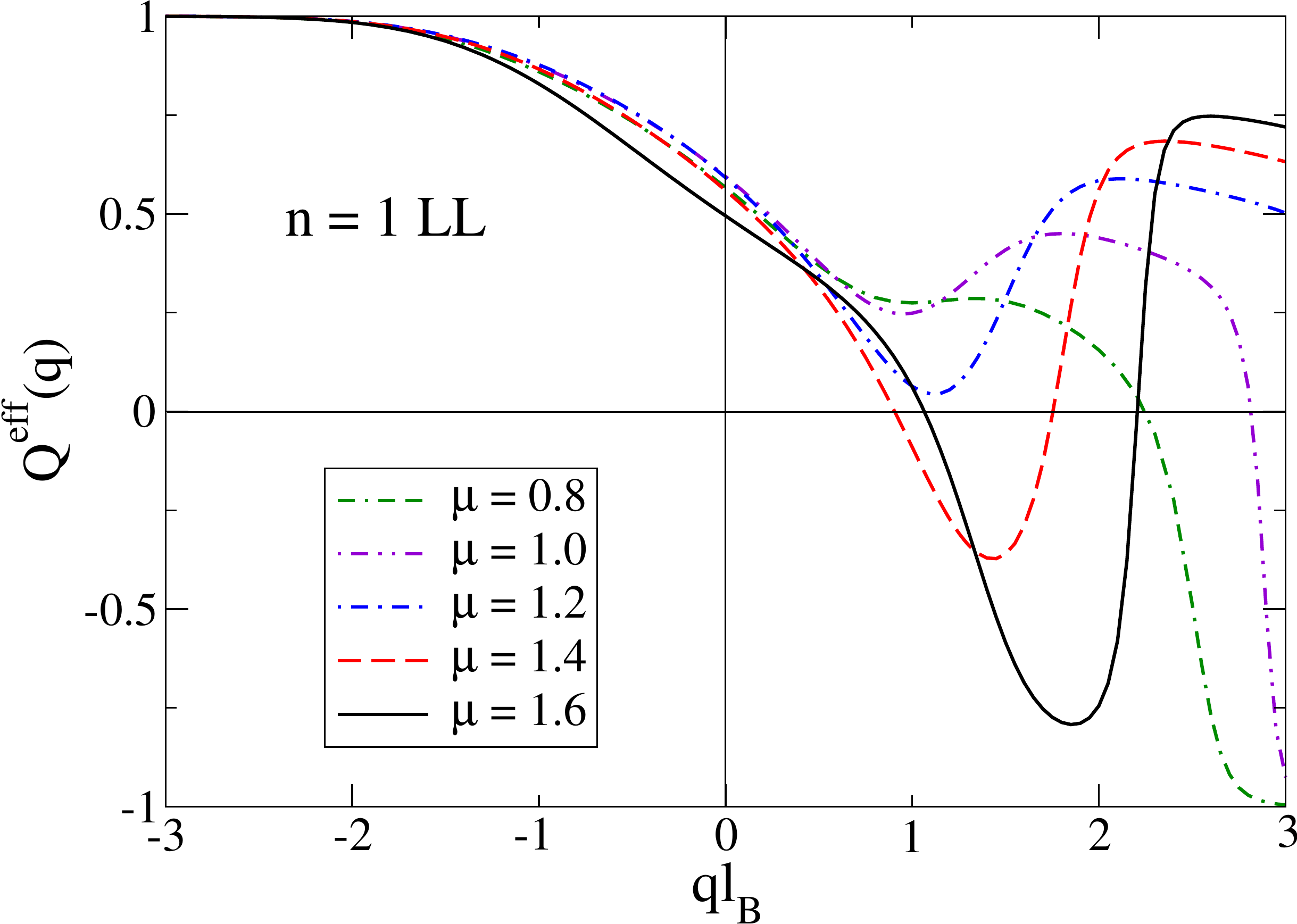}
\caption{\label{fig:chargen1} (Colour online) Effective Nambu charge
  $Q^{\text{eff}}(q)$ calculated using eq.~(\ref{eq:qeff}) for different
  values of $\mu$ (expressed in units of $\hbar v_F/l_B$), for the
  first ($n=1$) LL.}
\end{figure}

The Andreev edge states discussed here are superpositions of electron
and Nambu-hole Landau-level states. As $\mu$ increases, neighbouring LLs
get close in energy at a finite non-zero value of $q$ (as shown in
fig.~\ref{fig:dispersion}) and interact.  This interaction (level
repulsion) between neighboring LLs involves states with opposite sign
of charge.  Level repulsion between these states hybridises them
resulting in \textit{neutral} Andreev edge modes at that particular
value of $q$. This hybridisation always results in a pair of neutral
edge modes, one in each LL participating in the level repulsion. To
illustrate this further we show the numerically calculated
effective Nambu charge $Q^{\text{eff}}(q)$ for the first ($n=1$) LL in
fig.~\ref{fig:chargen1} for different values of $\mu$.  Comparing the
vanishing of the effective charge for $\mu=1.4\hbar v_F/l_B$ and
$\mu=1.6\hbar v_F/ l_B$ in Figs.~\ref{fig:effCharge} and
~\ref{fig:chargen1}, we can see that the level repulsion between the
zeroth and the first LLs results in neutral edge modes in each of them
at the same value of the guiding-center coordinate.

Our results can be easily extended to a NS junction based on graphene,
where each edge mode will have a twofold valley degeneracy assuming
that the boundary conditions at the NS interface does not mix the two
valleys~\cite{akh07}. As shown in ref.~\cite{par13}, the chiral
Majorana mode existing within the zeroth LL at $q=0$ for TIs does not
exist for graphene. However, we expect the neutral fermionic edge mode
predicted here (for $q\neq0$) to survive in graphene. 
Recently, some progress has been made towards the experimental
realisation of such heterostructures on graphene NS
junctions~\cite{ric12}.  

The hybridisation of electron-like and hole-like quantum Hall edge
channels was experimentally studied recently in InAs/GaSb ambipolar
quantum wells~\cite{nic13}. In these devices the hybridisation results
in an energy gap over the whole sample width, which can be observed in
transport measurements. Although similar in spirit, our system is
quite different from the one studied in ref.~\cite{nic13}. In
our NS junction the coupling between the electrons and the Nambu holes
is provided by the superconducting interface via Andreev
reflection. The superconducting region is therefore a necessary
ingredient for the presence of the neutral edge modes discussed in
this article. Moreover in our case the level repulsion \textit{does
  not} open an energy gap in the spectrum.

Strictly speaking our calculation is only applicable to magnetic
fields below the first critical field of the superconductor
($B\lesssim B_{c1}$).  For fields just above $B_{c1}$ a small but
finite number of vortices could change the properties of the edge
modes discussed here if these vortices are located close to the edge
(within distances of the order of the superconducting coherence
length). The effect of these vortices can be reduced by pinning the
vortices away from the edge. The order parameter induced in the left
half ($x<0$) by the superconductor on top of the sample will also lead
to a finite order parameter in the right half $(x>0)$ which will decay
exponentially away from the interface ($x=0$). However, the magnetic
field applied to the right side will quench this weak induced order
parameter since it is not supported by a genuine pairing interaction
(see, {\it e.g.}, ref.~\cite{belzig_dia} and references therein). As a
result of this residual order parameter, or due to a suppression of
the order parameter in the superconductor by the applied field, the
actual NS interface may be slightly shifted from $x=0$.  However, our
conclusions do not depend on the exact geometric location of the
electronic interface.

One way to investigate the effective charge of the edge modes
discussed here would be to generate an edge magnetoplasmon wave packet
by using a voltage pulse. The shape of this wave packet would
crucially depend on the effective charge of the fundamental density
excitations. Furthermore, even though charge will not be transported
by the neutral mode, heat will. 
This is because the group velocity $v_g = \partial\varepsilon_n(q)/
\partial(\hbar q)$ does not vanish at the value of $q$ at which $Q^{eff}=0$.
(For example, the neutral mode occurring in the $n=0$ Landau level at
$ql_B \approx 2.2$ has $v_g = 0.53\, v_F$, which corresponds to 79\%
of the chiral Majorana mode's velocity.)
Thus, by comparing heat and charge transport experiments, it should be
possible to establish the vanishing charge of these excitations.

\section{Conclusion}\label{sec:summary}
We have investigated the spectrum of dispersive edge states in a
superconductor -- topological-insulator junction in the presence of a
perpendicular magnetic field. The spectrum can change dramatically if
the chemical potential $\mu$ of the superconductor is raised above
some critical value $\mu_c$.  We show that the zeroth Landau level
supports additional \textit{neutral} edge modes at finite values of
the guiding-center coordinate when $\mu>\mu_c$, along with a neutral
Majorana mode at $q=0$. By investigating the energy spectra for
$\mu>\mu_c$ we find that these neutral modes are related to level
repulsion between the zeroth LL and the first ($n=\pm1$) LLs. We also
find that there are neutral modes at finite positive and negative
values of the guiding-center coordinate in $n=1$ and $n=-1$ LLs
respectively, corresponding to the same values of $q$ at which there
are neutral edge modes in the zeroth LL.  By calculating the effective
Nambu charge and its derivative we have determined the critical
chemical potential $\mu_c$ above which these neutral edge modes exist.
They arise due to the hybridisation of electron-like and hole-like LLs
involved in the level repulsion.  An experimental realisation of the
heterostructures we have investigated will provide further insight
into the nature of exotic Andreev edge state in the presence of a
perpendicular magnetic field.

\begin{acknowledgments}
R.P.T. and C.B. acknowledge support by the Swiss NSF and the NCCR Quantum
Science and Technology. V.M.S. was supported by the Swiss NSF.
\end{acknowledgments}


\begin{thebibliography}{10}

\bibitem{Klitzing+:80}
  \Name{Klitzing K., Dorda G. \and Pepper M.}
  \REVIEW{Phys. Rev. Lett.}{45}{1980}{494}.
  
\bibitem{Girvin+Prange:90}
  \Name{Prange R. E. \and Girvin S. M.}
  \Book{The Quantum Hall Effect}
  \Publ{Springer-Verlag, New York}
  \Year{1990}.

\bibitem{Zhang+:05}
  \Name{Zhang Y., Tan Y.-W., Stormer H. L. \and Kim P.}
  \REVIEW{Nature (London)}{438}{2005}{201}.
  
\bibitem{Abanin+:06}
  \Name{Abanin D. A., Lee P. A. \and Levitov L. S.}
  \REVIEW{Phys. Rev. Lett.}{96}{2006}{176803}.
  
\bibitem{Murakami+:03}
  \Name{Murakami S., Nagaosa N. \and Zhang S.-C.}
  \REVIEW{Science}{301}{2003}{1348}.

\bibitem{Kane+Mele:05}
  \Name{Kane C. L. \and Mele E. J.}
  \REVIEW{Phys. Rev. Lett.}{95}{2005}{226801}.
  
\bibitem{Bernevig+:06}
  \Name{Bernevig B. A., Hughes T. L. \and Zhang S.-C.}
  \REVIEW{Science}{314}{2006}{1757}.

\bibitem{TIreviews1}
  \Name{Hasan M. Z. \and Kane C. L.}
  \REVIEW{Rev. Mod. Phys.}{82}{2010}{3045}.
  
\bibitem{TIreviews2}
  \Name{Qi X.-L. \and Zhang S.-C.}
  \REVIEW{Rev. Mod. Phys.}{83}{2011}{1057}.
  
\bibitem{Andreev:64}
  \Name{Andreev A. F.}
  \REVIEW{Sov. Phys. JETP}{19}{1964}{1228}.

\bibitem{Beenakker:06}
  \Name{Beenakker C. W. J.}
  \REVIEW{Phys. Rev. Lett.}{97}{2006}{067007}.

\bibitem{akh07}
  \Name{Akhmerov A. R. \and Beenakker C. W. J.}
  \REVIEW{Phys. Rev. Lett.}{98}{2007}{157003}.

\bibitem{ric12}
  \Name{Rickhaus P., Weiss M., Marot L. \and Sch\"onenberger C.}
  \REVIEW{Nano Letters}{12}{2012}{1942}.

\bibitem{tiw13}
  \Name{Tiwari R. P., Z\"ulicke U. \and Bruder C.}
  \REVIEW{Phys. Rev. Lett.}{110}{2013}{186805}.
  
\bibitem{fu08}
  \Name{Fu L. \and Kane C. L.}
  \REVIEW{Phys. Rev. Lett.}{100}{2008}{096407}.

\bibitem{deg89}
  \Name{de Gennes P. G.}
  \Book{Superconductivity of Metals and Alloys}
  \Publ{Addison-Wesley, Reading, MA}
  \Year{1989}.

\bibitem{kiv90}
  \Name{Kivelson S. A. \and Rokhsar D. S.}
  \REVIEW{Phys. Rev. B}{41}{1990}{11693}.

\bibitem{nay01}
  \Name{Nayak C., Shtengel K., Orgad D., Fisher M. P. A. \and Girvin S. M.}
  \REVIEW{Phys. Rev. B}{64}{2011}{235113}.

\bibitem{fuj08}
  \Name{Fujita K., Grigorenko I., Lee J., Wang W., Zhu J. X., Davis J. C, Eisaki H., Uchida S. \and Balatsky A. V.}
  \REVIEW{Phys. Rev. B}{78}{2008}{054510}.

\bibitem{par12}
  \Name{Parameswaran S. A., Kivelson S. A., Shankar R., Sondhi S. L. \and Spivak B. Z.}
  \REVIEW{Phys. Rev. Lett.}{109}{2012}{237004}.

\bibitem{par13}
  \Name{Park S., Moore J. E. \and Sim H.-S.}
  \REVIEW{Phys. Rev. B}{89}{2014}{161408}.
  
\bibitem{nic13}
  \Name{Nichele F., Nath Pal A., Pietsch P., Ihn T., Ensslin K., Charpentier C. \and Wegscheider W.}
  \REVIEW{Phys. Rev. Lett.}{112}{2014}{036802}.

\bibitem{belzig_dia}
  \Name{Belzig W., Bruder C. \and Sch\"on G.}
  \REVIEW{Phys. Rev. B}{53}{1996}{5727}.
    
\end{thebibliography}
\end{document}